\documentclass[twocolumn,secnumarabic,amssymb, nobibnotes, aps, prd]{revtex4}
\usepackage{graphicx}

\begin{document}
\title{Investigation of possibility of creation of a superconductor quantum register.}
\author{A.A. Burlakov, V.L. Gurtovoi, S.V. Dubonos, A.V. Nikulov and V.A. Tulin}
\affiliation{Institute of Microelectronics Technology and High Purity Materials, Russian Academy of Sciences, 142432 Chernogolovka, Moscow District, RUSSIA.} 
\begin{abstract} Multiple and single measurements of quantum states of mesoscopic superconducting loops are carried out in order to investigate a possibility of macroscopic quantum superposition and of creation of a superconductor quantum register. Asymmetric superconducting rings are used in order to be convinced that single measurement gives result corresponding to one of permitted states and multiple one gives an average value on these states. We have measured magnetic dependencies of resistance, rectified voltage and critical current on these rings. The observed quantum oscillations of the resistance and the rectified voltage, corresponding to multiple measurement of quantum states, give evidence of two permitted states at the magnetic flux inside the ring divisible by half of the flux quantum. But the observed quantum oscillations of the critical current, corresponding to single measurement, not only does not confirm these two quantum states, but are in a direct contradiction with the observed oscillations of the resistance. It is assumed that the observed contradiction between the results of measurements made on the same ring can testify to violation of the principle of realism on the mesoscopic level that is a necessary condition for an opportunity of creation of a quantum computer. 
 \end{abstract}

\maketitle

\narrowtext

\section*{Introduction}

In hundred years after the publications of the first works which have founded bases of the quantum mechanics, a problem of its interpretation are again actual, as well as it was many years ago \cite{1}. The battles concerning interpretation of quantum mechanics stormed till 1930 years. Founders of the quantum theory, well understanding value of this problem, devoted a lot of time to discussions on this theme. After 1930 years the long period has followed during which majority of physicists did not give due attention to understanding of bases of quantum mechanics, having limited to use of the quantum formalism. This formalism was and remains surprisingly successful at the description of the various phenomena observed already not only in microcosm. However, the interpretation of the quantum mechanics remains as unclear as ever. In the book ``Character of Physical Laws'', Richard Feynman unabashedly declared, that "nobody understands quantum mechanics". But the wonderful difficulties of quantum mechanics are trivialized and the problem of its interpretation is swept aside as unimportant philosophical distractions by the bulk of the physical community.   

Many physics continue to think that it is not necessary to aspire to understand the quantum mechanics, if its formalism worked and continues to describe perfectly, how, at least, apparently, all observed phenomena. But some experts on quantum mechanics, in particular the Nobel prize winner on physics for 2003 of Anthon Leggett \cite{2}, specify that just now interpretation of quantum mechanics has ceased to be only philosophical problem. Development of technology and experimental techniques have made possible an experimental research of problems which were a subject of the pure philosophical dispute between Einstein and Bohr. Moreover, these problems have got practical importance. The most known and a vivid example here is Einstein-Podolsky-Rosen paradox \cite{3} and the idea of quantum calculations connected to it. This idea can be described at use of the formalism of quantum mechanics, as is done in the majority of articles and books. But at such formalistic approach it is frequently lost that quantum superposition of states and entanglement, underlying ideas of quantum computation, break the principle of realism. The contradiction between quantum superposition and the principle of local realism has been shown first by Einstein, Podolsky and Rosen as far back as in 1935 \cite{3} and expressed in formulas by Bell in 1964 \cite{4}. But no all researchers realize for the present profundity and importance of this contradiction for quantum physics. Because of the formalistic approach many researchers omit from the circumstance, important for search of ways of creation of a quantum computer, that it is necessary to break the principle of realism for realization of the idea of quantum computation. 

By the present time there are experimental evidences of violation of realistic prediction only at the level of elementary particles \cite{5}. Contrary to the belief of Einstein, Podolsky and Rosen \cite{3}, experiments give evidence correlation (Einstein-Podolsky-Rosen correlation) between results of measurements of photon states divided more than macroscopic distance. This experimental result disprove the principle of local realism, proceeding from which, authors \cite{3} have undertaken attack to the Copenhagen interpretation of quantum mechanics. But development of technology, even at the most optimistic variant, will not allow in the nearest years to create really working quantum computer at the atomic level. Therefore an opportunity of creation in foreseeable by the future really working quantum computer it is directly connected to a fundamental problem of existence of the EPR correlation at levels above atomic, first of all at mesoscopic level. The technology already for a long time has mastered the level micron sizes and now creation of nano-structures is not difficult problem. Therefore a proof of the entanglement existence at this level of sizes would point out a real way to creation of a quantum computer. 

But on the levels above  atomic, the quantum mechanics comes in the contradiction with principles not only local, but also macroscopic realism \cite{6}. Therefore, the proof of violation of predictions of realistic theories should be here, at least, not less strict and unambiguous than it has been made in experiments with photons \cite{5}. Such proofs are not obtained for the present. Some authors, not understanding profundity and fundamental nature of the problem, claim on experimental evidence of quantum superposition of macroscopic state, although their results can in no way prove violation of the principle of realism. One may say that this problem was not investigated for the present. It is known and, certainly, it is proved, that there are quantum phenomena on mesoscopic and macroscopic levels. Almost all these phenomena are connected anyhow to the Bohr's quantization. But existence of the Bohr quantization yet does not mean automatically existence of superposition of states. The Bohr quantization does not contradict to the principle of realism and there is not necessary to ask ``Is the flux there when nobody look?'' \cite{6} or even ``Is the moon there when nobody look?'' \cite{7}. 

Base element of a quantum computer is quantum bit -- qubit, i.e. a quantum system which can be in superposition of two quantum states \cite{8,9}. According to the universally recognized points of view \cite{10}, a superconducting ring is quantum system with two permitted states with the same energy when the magnetic flux inside it $\Phi$ is divisible by half of the flux quantum $\Phi_{0} = \pi \hbar/e$. Therefore such ring interrupted by Josephson junctions is considered as possible quantum bit, flux qubit \cite{11}. There can be no doubt that superconductivity is macroscopic quantum phenomenon. All multitude of experimental results give evidence of this. The superconducting state is described by uniform wave function and between two points of a superconductor, separated with any distance, exists phase coherence. It allows to hope that a chain of superconducting rings can be used as a quantum register \cite{12}. Modern technology can make such structure and creation of the quantum register with enough big number of qubits could become a reality if superposition of states and EPR correlation exists not only on the level of elementary particles. 

Superposition of two quantum states, according to conventional logic, assumes existence of these two states. But the term ``existence'' can be hardly applied in quantum world because of violation of the principle of realism. According to the quantum doctrine, a measurement does not, in general, reveal a preexisting value of the measured property \cite{13}. On contrary, the outcome of a measurement is brought into being by the act of measurement itself. Both Einstein and Bohr were agreed that the quantum theory can provide only a prediction of a result of measurement and can not provide a full account of ``how nature did it'' \cite{1}. But Einstein considered it as a defect of the theory whereas Bohr stated that it is distinctive feature of the quantum world and all hope of attaining a unified picture of objective reality must be abandoned. The experimental evidences of violation of realistic predictions force to agree with Bohr that we can say on result of measurement only. According to this point of view, superposition of quantum states assumes that single measurement should give result corresponding to one of the permitted states and multiple measurement should give average value on the both permitted states. The results of such measurements, carried out on asymmetric superconducting rings, are the content of the present article. 

\bigskip

\section{Multiple and single measurements of quantum states of a superconducting ring.}

We did not put the task to research of a possibility of state superposition. We wanted to check up, that measurements give that they should give according to our supposition about quantum states of a superconducting ring. Therefore we used rings without Josephson junction. 

\subsection{Quantum states of a superconducting ring. }

The spectrum of the permitted states of a superconducting ring is discrete because of the Bohr quantization
$$\oint_{l} dl p = \oint_{l} dl (mv_{s} + 2eA) = m\oint_{l} dl v_{s} + 2e\Phi  = n2\pi \hbar \eqno{(1)}$$
Velocity circulation of superconducting pairs along a contour $l$
$$\oint_{l} dl v_{s} = \frac{2\pi \hbar}{m} (n -\frac{\Phi}{\Phi_{0}}) \eqno{(2)}$$
can not be equal to zero at magnetic flux inside this contour not divisible by the flux quantum. It is consequence of quantization (1) generalized momentum $mv +2eA$, equal to the sum of products of mass on the velocity $mv$ and a charge of pair $2e$ on a vector potential $A$. Here it is important to remember, that the vector potential $A$ is not gauge-invariant value: $A$ does not differ physically from $A + \bigtriangledown \psi $, if $\psi $ a simple function without singularity. The integral from gradient of such function $\bigtriangledown \psi $ along a closed path is equal to a zero $\oint_{l} dl \bigtriangledown \psi $. Therefore the integral from a vector potential along a closed path, equal to a magnetic flux inside this path $\oint_{l} dl A = \Phi $, is gauge -invariant value, in contrary to the integral along no closed path. This feature is important for understanding of qualitative difference of superconducting loops without and with Josephson junction. 

According to (2) and to a condition of a constancy of value of superconducting current $I_{p} = sj_{p} = s2en_{s}v_{s}$ along ring circumference, the quantum number $n$ is unequivocally connected to gauge-invariant values $\Phi$ and $I_{p}$ which can be measured. But it becomes not so if the superconducting contour "is broken off" by Josephson junction, the current through which is proportional to sine $I_{p} = I_{c}\sin \Delta \phi$ of a phase difference $\Delta \phi$ of wave function \cite{14}. In this case the condition of quantization  (1) $\oint_{l} dl p = \oint_{l} dl \hbar \bigtriangledown \psi = n2\pi \hbar$ and the continuity of current $I_{p} = sj_{p} = s2en_{s}v_{s} = I_{c}\sin \Delta \phi$ do not give unequivocal connection between quantum number $n$ and the measured physical values $\Phi$ and $I_{p}$, since $\sin \Delta \phi= \sin (\Delta \phi+n2\pi )$. Nevertheless the current $I_{p}$ is not equal to zero at $\Phi \neq n\Phi_{0}$ in a superconducting loop with Josephson junction, as well as without it. In the case, considered in the majority of works, when a critical current of Josephson junction much less than a critical current of the loop $I_{p}  = -I_{c}\sin (2\pi \Phi /\Phi_{0})$ \cite{14}. 

The equilibrium direct current $I_{p}$, existing because of the Bohr quantization, has been named in the beginning of 60-years persistent current \cite{15}. Later this term began to be used for the equilibrium current observed in normal metals and semiconductors \cite{16}. Let note an important difference between a loop with and without Josephson junction: the persistent current equals zero $I_{p}  = -I_{c}\sin (2\pi \Phi /\Phi_{0}) = -I_{c}\sin (2\pi n + \pi ) = 0$ at full magnetic flux inside the loop  $\Phi = (n+0.5)\Phi_{0}$ in the first case whereas in the second case a superconducting state with $I_{p}= 0$ is forbidden at $\Phi = (n+0.5)\Phi_{0}$. It is more important that two permitted states with lowest energy and opposite directed velocities exist in second case because of the difference of the quantum number $n$ and $n+1$: $v_{s} \propto  n - \Phi /\Phi _{0} = 1/2$ at$ n+1$ and $-1/2$ at $n$.  In contrast to this a loop with Josephson junction can be two-state quantum system, considered as possible qubit \cite{11}, only due to the difference of full flux $\Phi = \Phi_{ext} + \Phi_{I}$  from the one $\Phi _{ext} = BS$ created by external magnetic field $B$ in the loop with area $S$: $I_{p}  = -I_{c}\sin (2\pi \Phi /\Phi_{0}) = -I_{c}\sin (2\pi \Phi_{I} /\Phi_{0} + \pi ) = I_{c}\sin (2\pi \Phi_{I} /\Phi_{0})$ at $\Phi_{ext} = (n+0.5)\Phi_{0}$. Two state can exist when the flux created by the persistent current $\Phi_{Ip} = LI_{p}$ is great enough, i.e. when parameter $\beta _{e} = 2\pi LI_{c}/\Phi_{0} > 1$  \cite{14}. Just for this case a possibility of superposition of macroscopic quantum states is assumed \cite{11} and its contradiction with of macroscopic realism is considered \cite{6}. But for an observation of two states assumed at $\Phi = (n+0.5)\Phi_{0}$, it is better to use a ring without Josephson junction since there is much more probable, than assumed quantum tunneling \cite{17}, a switching of a loop with Josephson junction between two permitted states by an uncontrollable noise which are at any real measurements. In this case any measurement will give an average value of the permitted states. 

\begin{figure}
\includegraphics{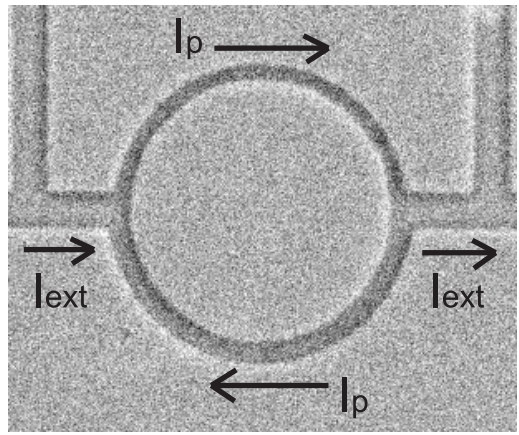}
\caption{\label{fig:epsart}  An SEM image of the asymmetric Al ring with radius $r = 2 \ \mu m$ and semi-ring width $w_{n} = 0.2 \ \mu m$, $w_{w} = 0.3 \ \mu m$. Directions accepted as positive of the external current $I_{ext}$ and the persistent current $I_{p}$ are shown by arrows.}
\end{figure}

\begin{figure}
\includegraphics{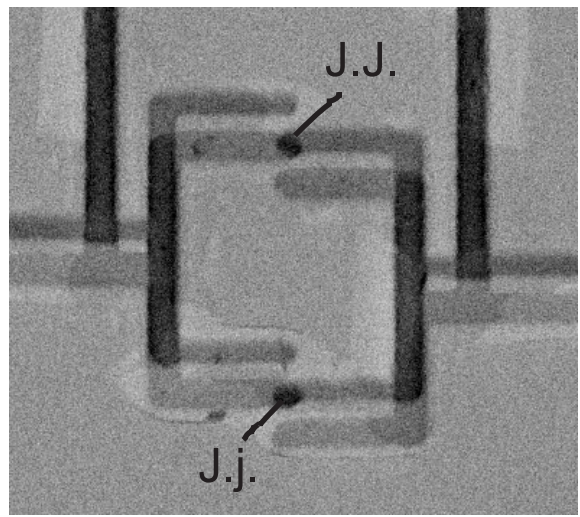}
\caption{\label{fig:epsart}   An SEM image of the asymmetric Al superconducting quantum interferometer, i.e. a superconducting loop with two Josephson junction (J.j.),  fabricated by us with suspended shadow mask technique. The size of the square loop  is equal $4 \ \mu m$.}
\end{figure}

In a ring without Josephson junction, Fig. 1, where the quantum number $n$ is unequivocally connected to the real physical values $\Phi$ and $I_{p}$, transition between quantum states is possible only at break of connectivity of wave function. Therefore uncontrollable noise should not influence on results of measurements. The energy difference between the permitted levels of any real superconducting ring, proportional to number of pairs in the ring, is much greater than energy of thermal fluctuations $k_{B}T$, \cite{18}. Therefore the level with the least energy, i.e. with the least square of velocity $v_{s}^{2} \propto  (n - \Phi /\Phi _{0})^{2}$, has overwhelming probability. Numerous observations \cite{19} of the Little-Parks oscillations \cite{20} of resistance of superconducting loop give experimental evidence this overwhelming probability even near the critical temperature $T \approx T_{c}$. According to the universally recognized explanation \cite{10}, the periodic change of resistance $R(\Phi /\Phi _{0})$ of a thin-walled superconducting cylinder \cite{20} or ring \cite{19}, measured at the temperature corresponding to resistive transition $R_{ln} > R_{l}> 0$, is result of critical temperature change $\Delta R(\Phi /\Phi _{0}) \propto - \Delta T_{c} (\Phi /\Phi _{0})$. The $T_{c}$ reduction  at $\Phi \neq n\Phi_{0}$ is connected with increase in energy of superconducting state $\propto v_{s}^{2} (\Phi /\Phi _{0})$ at $v_{s} \propto   n - \Phi /\Phi _{0} \neq 0$: $-\Delta T_{c} (\Phi /\Phi _{0}) \propto v_{s}^{2} \propto  (n - \Phi /\Phi _{0})^{2}$  \cite{10}. Experimental observation of periodic reduction of critical temperature of a ring with magnetic field \cite{19} corresponds enough well to $\Delta T_{c} (\Phi /\Phi _{0}) \propto  -(n - \Phi /\Phi _{0})^{2}$, where $n$ is the integer corresponding to minimum of $(n - \Phi /\Phi _{0})^{2}$. 

All observations of the Little-Parks oscillation testify that the square velocity of superconducting pairs has maximal values at $\Phi = (n+0.5)\Phi_{0}$ and minimal at $\Phi = n\Phi_{0}$. Proceeding from the universally recognized explanation \cite{10} it is possible to assume also, that velocity changes sign at $\Phi = (n+0.5)\Phi_{0}$. But, strictly speaking, the Little-Parks experiment cannot prove the direction change, since the resistance variation $\Delta R \propto v_{s}^{2}(\Phi/\Phi_{0})$. A change of a phases difference between two rings connected by Josephson junctions was interpreted in \cite{21} as a consequence of the direction inversion of velocity of superconducting pairs in one of the rings. In the present work we put a task to fix this inversion, measuring the critical current of the asymmetric ring, as shown on Fig. 1. 

\subsection{Critical current of an asymmetric ring.}

The critical current both superconducting quantum interferometer \cite{10}, i.e. a superconducting loop with two Josephson junction, Fig. 2, and a ring \cite{22}, is periodic function of magnetic field. The critical current of an asymmetric ring, Fig. 1, in contrast to symmetric one, should depend not only on value, but also on the direction of velocity of superconducting pairs. The velocity $v_{sn}$ in narrow (with section $s_{n}$) and $v_{sw}$ in wide (with section $s_{w}$) semi-rings are determined by the value of the measuring current $I_{ext } = I_{n} + I_{w} = s_{n}j_{n} + s_{w}j_{w} = 2en_{s} (s_{n}v_{sn} + s_{w}v_{sw})$ and the Bohr's quantization (1), according to which (2) $l_{n}v_{sn} - l_{w}v_{sw} = l (v_{sn} - v_{sw})/2 = (2\pi \hbar /m) (n - \Phi /\Phi _{0})$: $v_{sn} = I_{ext}/2en_{s} (s_{n}+s_{w}) + (2\hbar /mr)s_{w}/[(s_{n}+s_{w})] (n -  \Phi /\Phi _{0})$ and $v_{sw} = I_{ext}/2en_{s} (s_{n}+s_{w}) - (2\hbar /mr)s_{n}/[(s_{n}+s_{w})] (n -  \Phi /\Phi _{0})$. Here and further left-right direction for $I_{ext}$, $v_{sn}$, $v_{sw}$ and clockwise direction for $I_{p}$ are chosen as positive, Fig. 1. Transition in the resistive state occurs when pair velocity reaches critical value in narrow $\vert v_{sn}\vert  = v_{sc}$ or wide $\vert v_{sw}\vert  = v_{sc}$ semi-ring. At the positive (left-right ) direction of the measuring current $I_{ext}$ it takes place in narrow semi-ring when the value $I_{p} \propto n -  \Phi /\Phi _{0}$ is positive (has clockwise direction) and in the wide one when $I_{p} \propto n -  \Phi /\Phi _{0}$ is negative. The critical current is equal in the first case 

\begin{figure}
\includegraphics{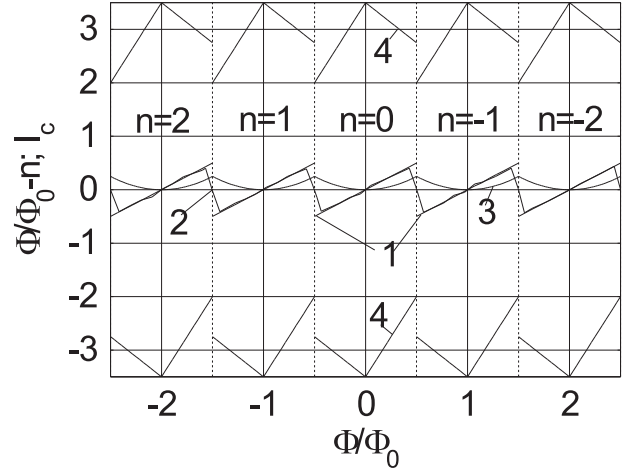}
\caption{\label{fig:epsart}   Magnetic dependencies of the lowest permitted velocity $v \propto n -\Phi/\Phi_{0}$ (1), the average equilibrium velocity $\overline{v} \propto \overline{n} -\Phi/\Phi_{0}$ (2), the velocity square $v^{2} \propto (n -\Phi/\Phi_{0})^{2}$ (3) and the critical current  $I_{c+}, I_{c-}$ (4) expected according to the equations (3a) and (3b) for a ring with $s_{w}/s_{n} = 2$,  $I_{c0} = 3.5 \ \mu A$, $I_{p,A} = 0.5  \ \mu A$.}
\end{figure}

$$I_{c+} = I_{c0} - 2I_{p,A}\vert n - \frac{\Phi }{\Phi _{0}}\vert (1 + \frac{s_{w}}{s_{n}})  \eqno{(3a)} $$
and in the second case
$$I_{c+} = I_{c0} - 2I_{p,A}\vert n - \frac{\Phi }{\Phi _{0}}\vert (1 + \frac{s_{n}}{s_{w}})  \eqno{(3b)} $$

Here $I_{c0} = 2en_{s}(s_{n}+s_{w}) v_{sc}$  is the critical current at $\Phi /\Phi _{0} = n$; $I_{p,A} = 2en_{s}(\hbar /mr)s_{n}s_{w}/(s_{n}+s_{w}) $ is the value corresponding to amplitude of the persistent current oscillations when the $n -  \Phi /\Phi _{0}$ value changes between $-0.5$ and $0.5$. For the critical current $I_{c-}$ measured in the negative (right-left) direction, expressions are similar but the (3a) is applied at $I_{p} \propto n -  \Phi /\Phi _{0} < 0$ and (3b) at $I_{p} \propto n -  \Phi /\Phi _{0} > 0$. The expected dependencies $I_{c+} (\Phi /\Phi _{0})$, $I_{c-} (\Phi /\Phi _{0})$ are plotted on Fig. 3 for values $s_{w}/s_{n} = 2$, $I_{c0} = 3.5 \ \mu A$, $I_{p,A} = 0.5 \ \mu A$. 

A break should be expected in the dependencies $I_{c+} (\Phi /\Phi _{0})$ , $I_{c-} (\Phi /\Phi _{0})$ (3), Fig.3, because of the direction inversion of the persistent current at $\Phi = (n+0.5)\Phi _{0}$. The difference (anisotropy) of the critical currents measured in opposite directions, should be proportional to the persistent current $I_{c,an } = I_{c+} - I_{c-} = I_{p} (s_{w}/s_{n} - s_{n}/s_{w}) = 2I_{p, A} (n - \Phi/\Phi_{0})(s_{w}/s_{n} - s_{n}/s_{w})$ according to (3), that is should be periodic, sign-variable function of magnetic field $I_{c,an} (\Phi /\Phi _{0})$. This anisotropy of the current-voltage curves should result in the rectification of an alternating current, for example $I_{ext} = I_{0} \sin  (2\pi ft)$. The rectified voltage $V_{dc} = \Theta ^{-1}\int_{\Theta }dt V(I_{ext}(t))$ should be periodic sign-variable function of magnetic field $V_{dc}(\Phi /\Phi _{0})$. Such quantum oscillations of the rectified voltage were observed on an asymmetric aluminum ring \cite{23} and superconducting quantum interferometer \cite{24}. 

\subsection{Multiple measurements of quantum states.}

Measurement of the rectified voltage \cite{23} corresponds multiply measurement of quantum states of the superconducting ring. We have carried out detailed measurements of magnetic dependencies of the rectified voltage $V_{dc}(\Phi /\Phi _{0})$ both on single rings, with various asymmetry $s_{w}/s_{n} = 1.25 \div  2$, and systems of rings with the identical diameter, connected in series. The investigations were carried out on aluminum nano-structures at temperatures $T = 1.19 \div  1.3 \ K$. Film structures with temperature of superconducting transition $T_{c} = 1.23 \div 1.27 \ K$, the resistance on square $\approx 0.5 \ \Omega /\diamond $ at $T = 4.2 \ K$ and the resistance relation $R (300K)/R (4.2K) = 3$ were used. All rings had identical diameter $2r = 4 \ \mu m$. Thickness of the film used for creation of structure was $d = 40 \div 50 \ nm$. We had been investigated rings with identical length $l_{w} = l_{n} =\pi r = 6.3 \ \mu m$ and different width of semi-rings: $w_{w} = 0.4 \ \mu m$, $w_{n} = 0.2 \ \mu m$; $w_{w} = 0.3 \ \mu m$, $w_{n} = 0.2 \ \mu m$; $w_{w} = 0.25 \ \mu m$, $w_{n} = 0.2 \ \mu m$, and also symmetric rings with $w_{w} = w_{n} = 0.4 \ \mu m$. Semi-rings sections $s_{w} =w_{w}d \approx  0.01 - 0.02 \ \mu m^{2}$ and $s_{n} =w_{n}d \approx  0.008 - 0.02 \ \mu m^{2}$ was less than square of correlation length $\xi ^{2} (T)$ and London penetration depth of magnetic field $\lambda _{L}^{2} (T)$ in all cases. Such superconducting channels with small section can be considered as one-dimensional. 

\begin{figure}
\includegraphics{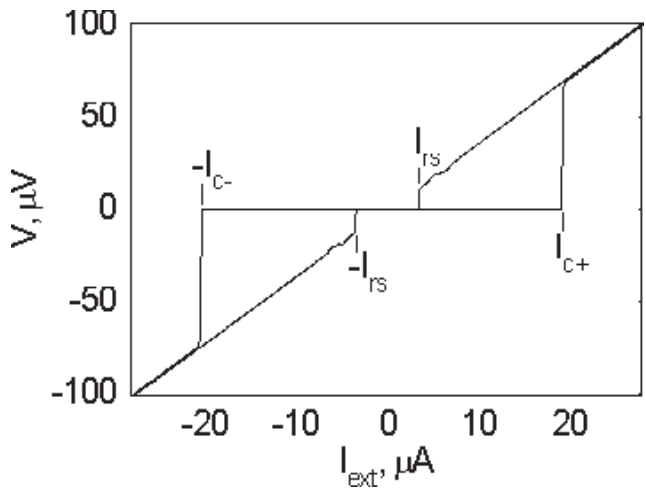}
\caption{\label{fig:epsart}   A typical current-voltage curve measured on an asymmetric aluminum ring.}
\end{figure}

Current-voltage curves (CVC) of such superconducting aluminum rings, Fig. 1, have a hysteresis and sharp transition in the resistive state at $Ò < 0.985T_{c}$, Fig. 4. The value of critical current $I_{c+}$, $I_{c-}$, at which this transition is observed, corresponds to the value of critical velocity $v_{sc}$ of superconducting pairs \cite{10}. It is typical for superconducting channel with small section. Our measurements of the CVC have shown that the value $I_{rs}$ of the current, at which returning in superconducting state is observed, Fig.4, does not depend on magnetic field and direction of the measuring current. In contrast to this the critical current $I_{c+}$, $I_{c-}$ and their difference (anisotropy) $I_{c,an} = I_{c+} - I_{c-}$ change periodically with value of magnetic field. The rectified voltage should be proportional to average value (obtained at multiple measurement) of the anisotropy of the critical current $V_{dc} \propto <I_{c,an}> = <I_{c+} - I_{c-}>$ at $I_{0} > I_{c+}, I_{c-}$, since it the voltage average on time $V_{dc} = \Theta ^{-1}\int_{\Theta }dt V(I_{ext}(t))$.

For investigation of magnetic dependencies of the rectified voltage and other properties we used the following technique. Magnetic field $B_{sol} = k_{sol}I_{sol}$, perpendicular planes of sample, was created by a current $I_{sol }$ in the copper solenoid located outside of cryostat. The gauge constant of the solenoid was equal $k_{sol} = 129 \ G/A$. The periods $B_{0} = \Phi _{0}/S = 1.4 \div 1.6 \ G$ of all periodic dependencies $V_{dc} (B)$, $R (B)$, $I_{c}(B)$ met to the area of the rings $S = \pi r^{2} = 14.8 \div  13.0 \ \mu m^{2}$, used for measurement. Where the $r = 2.2 \div  2.0 \ \mu m$ values are close to internal radius of the given ring. For reduction of the Earth magnetic field the cryostat was shielded. The residual field was $B_{res} \approx  0.15 \ G$, i.e. about one tenth of the period $B_{0}$ of observed oscillations. Owing to the incomplete shielding a minimum of resistance $R(B_{sol})$ and zero value of the rectified voltage $V_{dc} (B_{sol})$ were observed at $B_{sol} =-B_{res} \approx  -0.15 \ G$. The measured dependencies are plotted in the paper as function of the magnetic flux inside the ring $\Phi  = SB = S (B_{sol} + B_{res})$, created by full external field $B_{sol} + B_{res}$. Exact value of the ring area $S$ got out of the condition of equality of the period oscillations $V_{dc} (B)$, $R (B)$, $I_{c}(B)$ to the flux quantum $S = \Phi_{0}/B_{0}$. Exact value of the residual magnetic field $B_{res}$ got out of the condition of minimum $R (\Phi )$ and $V_{dc}(\Phi ) = 0$ at $\Phi = 0$, and the condition $I_{c-}(\Phi )= I_{c+}(-\Phi )$. The value $B_{res}$ was approximately identical in all cases. We used approximation $\Phi  = \Phi_{ext} + \Phi_{I} \approx \Phi_{ext} = S (B_{sol} +B_{res})$, i.e. neglected the magnetic flux $\Phi_{p}$, created by the external current $I_{ext}$ and the persistent current $I_{p}$, because of its negligible value $\Phi_{I} < 0.04\Phi _{0}$ at the small values of the ring inductance $L = 2 \ 10^{-11} \ H$, $I_{ext}$ and $I_{p}$ in our work.

\begin{figure}
\includegraphics{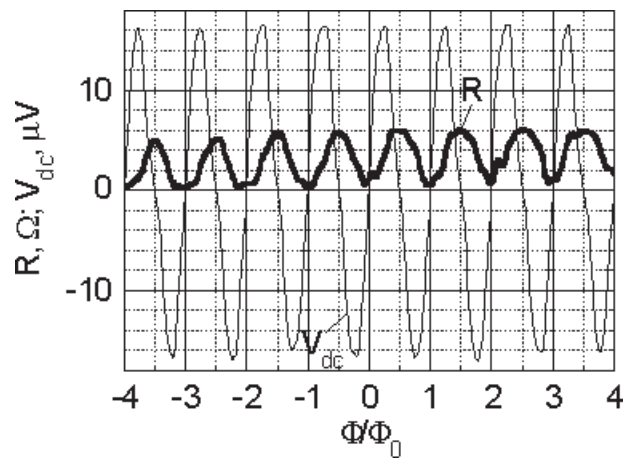}
\caption{\label{fig:epsart}   The quantum oscillations of the rectified voltage $V_{dc}$, induced by the ac current with frequency f = 0.5 kHz and amplitude $I_{0} = 20 \ \mu A$ at $T = 0.973 T_{c}$, and the Little-Parks oscillations $R$ at $T = 0.998 T_{c}$, measured on the ring with $w_{n} = 0.2 \ \mu m$,  $w_{w} = 0.3 \ \mu m$ and $T_{c} = 1.288 \ K$.}
\end{figure}

Our measurements have shown that the rectified voltage $V_{dc}(\Phi /\Phi _{0})$ appears when the amplitude of external current $I_{0}$ exceeds minimal of the critical values $I_{0} > min (I_{c+}, I_{c-})$, its amplitude reaches a maximum at $min (I_{c+}, I_{c-}) <I_{0} < max(I_{c+}, I_{c-})$ and decreases with further increase $I_{0}$. Such behaviour is observed for all 
interval of frequencies (from $f = 10 \ Hz$ to $f = 1 \ MHz$) of the external current and temperature ($T = 0.95 \div  0.995 T_{c}$), investigated by us. The quantum oscillations of the rectified voltage $V_{dc}(\Phi /\Phi _{0})$ observed on all rings cross zero value at $\Phi  = n\Phi _{0}$ and $\Phi  = (n+0.5)\Phi _{0}$, Fig. 5, in all cases. 

Our observations of these quantum oscillations confirm periodic change of sign and value of equilibrium velocity of superconducting pairs determined by the quantization (1). It is important to emphasize that the rectified voltage which should correspond to multiple measurements of the velocity along the ring circumference of $V_{dc} \propto  <I_{c,an}> = <I_{p}>(s_{w}/s_{n} - s_{n}/s_{w}) = 2I_{p,A}(<n> - \Phi /\Phi _{0}) (s_{w}/s_{n} - s_{n}/s_{w}) \propto  <v_{s}>$ equals zero at $\Phi  = n\Phi _{0}$ and $\Phi  = (n+0.5)\Phi _{0}$. According to the condition of quantization, the average velocity $<v_{s}> \propto <n> - \Phi /\Phi _{0}$ equal to zero at $\Phi  = n\Phi _{0}$ because the single permitted state with the least energy has zero velocity $<v_{s}> =  v_{s} \propto n - \Phi /\Phi _{0} = n-n = 0$. In contrast to this, the state with zero velocity is forbidden at $\Phi  = (n+0.5)\Phi _{0}$. But it is assumed that there are two permitted states, $n$ and $n+1$, having equal and opposite directed velocity and identical probability. Therefore the average velocity is assumed to equal zero $<v_{s}>  \propto <n> - \Phi /\Phi _{0} \propto 0.5 + (-0.5) = 0$. Our observations $V_{dc} = 0$ at $\Phi  = (n+0.5)\Phi _{0}$ and $V_{dc} \neq 0$ at $\Phi  \neq (n+0.5)\Phi _{0}$, Fig. 5, corroborate this assumption. The energy of states $n$, $n+1$ and, hence, their probabilities are not equal at a deviation of $\Phi$ from $\Phi  = (n+0.5)\Phi _{0}$. Therefore the rectified voltage $V_{dc} \propto <v_{s}>$ has positive or negative value, depending on a sign of the deviation, Fig. 5.

Our measurements of the Little-Parks oscillations of resistance $R(\Phi/\Phi_{0})$ of the same asymmetric rings have shown that the $R(\Phi/\Phi_{0})$ minimums are observed at $\Phi  = n\Phi _{0}$ and the $R(\Phi/\Phi_{0})$ maximum are observed at $\Phi  = (n+0.5)\Phi _{0}$, Fig. 5, as well as for symmetric rings. This result, together with the observation $V_{dc} = 0$ at $\Phi  = n\Phi _{0}$ and $\Phi  = (n+0.5)\Phi _{0}$, give evidence single permitted state at $\Phi  = n\Phi _{0}$ and two permitted state at $\Phi  = (n+0.5)\Phi _{0}$ since $\Delta R(\Phi/\Phi_{0}) \propto  v_{s}^{2}$, according to the universally recognized explanation \cite{10}. It is impossible logically to explain the observation of the maximum of $\Delta R(\Phi/\Phi_{0}) \propto  v_{s}^{2}$ and $V_{dc} \propto <v_{s}> = 0$ without the assumption of two permitted states with the same probability at $\Phi  = (n+0.5)\Phi _{0}$. 

The measurements of the Little-Parks oscillations $R(\Phi/\Phi_{0})$ are made at the temperatures corresponding to superconducting transition where thermal fluctuations switch a ring between states with different connectivity of wave function \cite{18} and the permitted states $n$, $n+1$. Therefore these measurements correspond to multiple measurements of the square velocity of pairs as well as the measurements of the rectified voltage $V_{dc}(\Phi /\Phi _{0})$ correspond to multiple measurements of the pair velocity. Thus, we may state, that multiple measurements give evidence of two permitted states $n$, $ n+1$ at $\Phi  = (n+0.5)\Phi _{0}$ with velocities equal on value and opposite on direction $v_{s} \propto n - \Phi /\Phi _{0} = 0.5$ or $-0.5$. It is necessary to confirm this result by measurement of the critical current of the same asymmetric ring. This measurement should correspond to single measurement of quantum states. 

\subsection{Single measurement of quantum states.} 

Our measurements have shown, that at achievement by measuring current $I_{ext}$ of a critical value $I_{c+}$, $I_{c-}$ whole structure turns by jump into the resistive state, Fig.4. Such sharp transition, observed both for single ring and for system of 20 rings with the common length $160 \ \mu m$, is typical for narrow superconducting channel. The jump in the normal state $n_{s} = 0$ of any segment of structure at the critical velocity of pairs $v_{s} = v_{sc}$ induces, because of the proximity effect, a reduction of pair density $n_{s}$ in adjacent segments. This reduction results to increase in velocity $v_{s}$ in these segment up to critical value and to transition in the normal state since the current $I = s2en_{s}v_{s}$. The energy difference between the permitted levels in a ring remains very high up to its transition in the resistive state  since the density of pairs decreases not strongly at the increase of their velocity up to the critical value $n_{s}(v_{s}=v_{sñ}) = 2n_{s}(v_{s}=0)/3$ \cite{10}. Therefore, any switching of the ring between the permitted states by thermal fluctuations or even uncontrollable noise is impossible. A ring falls down in a superconducting state with one of values of quantum number $ n$, when a measuring current $I_{ext}$ decreases down to $I_{rs}$, Fig. 4, and should remain in this quantum state up to the transition in the resistive state at $I_{c+}$ or $I_{c-}$. Therefore the measurement of critical currents $I_{c+}$, $I_{c-}$ should correspond to single measurement of quantum states of the ring: the value $I_{c+}$, $I_{c-}$ measured at $\Phi  = (n+0.5)\Phi _{0}$ should correspond to a state $n$ or $n+1$, but no their average value. In contrast to this it is impossible to guarantee that a loop with Josephson junction is not switched between the quantum states $n$, $n+1$ by an uncontrollable noise. It is advantage of the ring without Josephson junction, Fig. 1, over the loop with Josephson junctions, Fig. 2, for the single measurement of two states at $\Phi  = (n+0.5)\Phi _{0}$.

For the single measurement of quantum states we used the following technique. Dependence of the critical current on magnetic field $I_{c+}(B)$, $I_{c-}(B)$ was measured from periodically repeating CVC ($10 \ Hz$) in slowly varying magnetic field ($\sim 0.01 \ Hz$) on the following algorithm: the condition of a finding of the structure in superconducting state was checked, and then, when the measured voltage excesses a threshold value, the magnetic field and the critical current were measured. Thus the critical current in positive $I_{c+}(B)$ and negative $I_{c-}(B)$ directions of the measuring current $I_{ext}$ was consistently measured. For measurement of one dependence $I_{c+}(B)$, $I_{c-}(B)$, containing 1000 points, it was required about 100 seconds.

\begin{figure}
\includegraphics{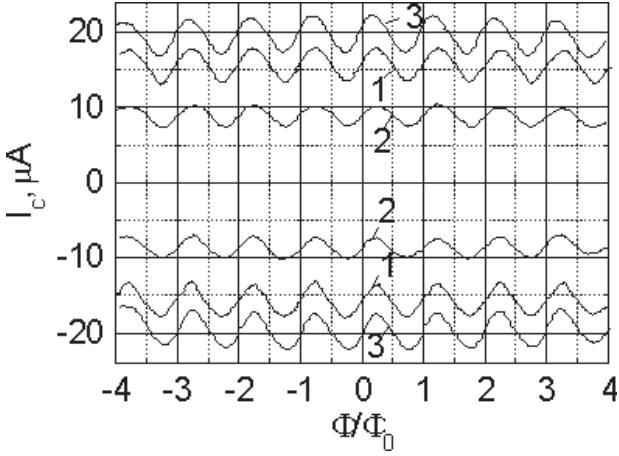}
\caption{\label{fig:epsart}  The quantum oscillations of the critical current $I_{c+}$, $I_{c-}$ measured in opposite directions on the rings with different asymmetry: 1) $w_{n} = 0.2 \ \mu m$,  $w_{w} = 0.25 \ \mu m$ at $T = 0.975 T_{c}$; 2) $w_{n} = 0.2 \ \mu m$,  $w_{w} = 0.3 \ \mu m$ at $T = 0.987 T_{c}$; 3) $w_{n} = 0.2 \ \mu m$,  $w_{w} = 0.4 \ \mu m$ at $T = 0.969 T_{c}$.}
\end{figure}

According to the conventional logic if mutiple measurements have proved existence of two states $n$, $n+1$ at $\Phi  = (n+0.5)\Phi _{0}$, single measurement should confirm it. It means, that the break should be observed on the $I_{c+}(\Phi /\Phi_{0})$, $I_{c-}(\Phi /\Phi_{0})$ dependencies of asymmetric rings at $\Phi  = (n+0.5)\Phi _{0}$, Fig. 3. However our measurements have not revealed such breaks. Moreover, we have received the oscillations $I_{c+}(\Phi /\Phi_{0})$, $I_{c-}(\Phi /\Phi_{0})$, Fig. 6, qualitatively distinguished from the one (3), Fig. 3, expected according the condition of quantization (2). We have found out, that the oscillations of the critical current measured in opposite directions, are similar $I_{c-}(\Phi/\Phi_{0}) = I_{c+}(\Phi/\Phi_{0}+ \Delta \phi)$, i.e. are superposed at a shift of one of the dependencies on the value equal to half of the flux quantum $\Delta \phi = 0.5$. We had been measured the dependencies $I_{c+}(\Phi /\Phi_{0})$, $I_{c-}(\Phi /\Phi_{0})$ on rings with various relation of semi-ring sections $s_{w}/s_{n} = w_{w}/w_{n} = 1.25; 1.5; 2$, i.e. with various anisotropy, in an interval of temperatures $T = 0.96 \div 0.99 T_{c}$ in which the value of a critical current changed on the order, from $3 \ \mu A$ to $30 \ \mu A$. Measurements have shown that the shift equals to half of the flux quantum with accuracy 0.02, $\Delta \phi = 0.5 \pm 0.02$, in all these cases, in spite of the difference of the anisotropy and the critical current. The observations of many periods of the $I_{c+}(\Phi /\Phi_{0})$, $I_{c-}(\Phi /\Phi_{0})$ oscillations, up to 30, allow us to determine the shift with this high accuracy. 

It is obvious that the CVC of an absolutely symmetric ring should be symmetric $I_{c-}(\Phi /\Phi_{0}) = I_{c+}(\Phi /\Phi_{0})$. The measurements, which have been carried out by us on rings with identical semi-rings width $w_{w} = w_{n} = 0.4 \ \mu m$, have shown what even small uncontrollable asymmetry results in a small shift of the $I_{c+}(\Phi /\Phi_{0})$, $I_{c-}(\Phi /\Phi_{0})$ dependencies. The best coincidence of the oscillations $I_{c+}(\Phi /\Phi_{0})$, $I_{c-}(\Phi /\Phi_{0})$ measured by us on two rings with $w_{w} = w_{n} = 0.4 \ \mu m$ is observed at $\Delta \phi \approx  0.05$ and $\Delta \phi \approx 0.07$. We have measured also the oscillations $I_{c+}(\Phi /\Phi_{0})$, $I_{c-}(\Phi /\Phi_{0})$ on a ring with $w_{w} = w_{n} = 0.4 \ \mu m$ along most length of the semi-rings $l_{w} = l_{n} =  6.3 \ \mu m$, but with the reduced width $0.3 \ \mu m$ of a small segment with length $0.5 \ \mu m$. This small controllable asymmetry increases the shift  up to $\Delta \phi \approx  0.25$. Thus, even very small asymmetry results in shift of the dependencies the critical current measured in opposite directions. 

\subsection{Conformity between the measurement results of the oscillations of the critical current and the rectified voltage.}

Although the observed, Fig. 6, and expected, Fig. 3, oscillations of the critical current in magnetic field of asymmetric rings are different in essence, in both cases there is a periodic change of anisotropy of the critical current $I_{c,an }(\Phi /\Phi_{0}) = I_{c+}(\Phi /\Phi_{0}) - I_{c-}(\Phi /\Phi_{0})$, explaining the observed periodic dependence of the rectified voltage $V_{dc}(\Phi /\Phi_{0})$. Our measurements have shown that maximums $V_{dc}(\Phi /\Phi_{0})$ correspond to minimums $I_{c+}(\Phi /\Phi_{0})$, $I_{c-}(\Phi /\Phi_{0})$ and $V_{dc}(\Phi /\Phi_{0}) \propto -I_{c,an }(\Phi /\Phi_{0})$, Fig. 7. The magnetic dependencies of the rectified voltage calculated from the observed CVC and the $I_{c+}(\Phi /\Phi_{0})$, $I_{c-}(\Phi /\Phi_{0})$ dependencies for different amplitude $I_{0}$ of the external current meet with the oscillations $V_{dc}(\Phi /\Phi_{0})$ induced by the ac current with appropriate amplitude $I_{0}$. 

\begin{figure}
\includegraphics{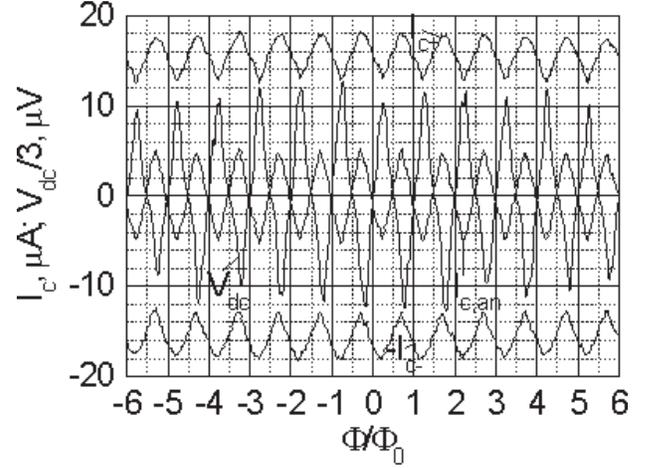}
\caption{\label{fig:epsart}   The quantum oscillations of the critical  current $I_{c+}$ $I_{c-}$,  its anisotropy  $I_{c,an} = I_{c+} - I_{c-}$ and the rectified voltage $V_{dc}$ measured on the rings with $w_{n} = 0.2 \ \mu m$,  $w_{w} = 0.25 \ \mu m$ at $T = 0.975 T_{c}$.}
\end{figure}

The observed oscillations of anisotropy of the critical current $I_{c,an }(\Phi /\Phi_{0})$ cross zero value at $\Phi  = n\Phi _{0}$ and $\Phi  = (n+0.5)\Phi _{0}$, as well as the rectified voltage $V_{dc}(\Phi /\Phi_{0})$. However it is impossible to draw a conclusion from this experimental fact that our measurements of the critical current correspond not single, but to multiple measurements. Here it is important to emphasize, that the reasons of the observed anisotropy of the critical current, Fig. 7, differs in essence from the reason of the expected anisotropy (3), Fig. 3. We expected that the anisotropy should result from a difference of the functions $I_{c+}(\Phi /\Phi_{0})$, $I_{c-}(\Phi /\Phi_{0})$, describing the oscillations of the critical current measured in opposite directions Fig.3. But we have found that in the observed oscillations of the critical current its anisotropy results from the difference of argument of these functions: $I_{c-}(\Phi/\Phi_{0}) = I_{c+}(\Phi/\Phi_{0}+ \Delta \phi) \neq I_{c+}(\Phi/\Phi_{0})$ at $\Delta \phi \neq 0$. It is very strange that no functions but their arguments change at the appearance of a ring asymmetry. The shift should be equal zero  $\Delta \phi = 0$  for the oscillations of the critical current measured on an absolutely symmetric ring since the CVC should be symmetric $I_{c-}(\Phi/\Phi_{0}) = I_{c+}(\Phi/\Phi_{0})$ in this case. And we observed that $I_{c-}(\Phi/\Phi_{0}) = I_{c+}(\Phi/\Phi_{0} + 0.05) \approx I_{c+}(\Phi/\Phi_{0}$ for the oscillations measured on the ring with $w_{w} = w_{n} = 0.4 \ \mu m$ closed to the symmetric one. But already for the ring with enough small asymmetry $w_{w}/w_{n} = 1.25$ the maximum shift $\Delta \phi = 0.5$ is observed, i.e. the argument of the functions $I_{c+}(\Phi /\Phi_{0})$, $I_{c-}(\Phi /\Phi_{0})$ changes on $\pm 0.25$ relatively the function describing the oscillations measured on the symmetric loop. This shift provides the rectification effect observed on asymmetric rings but it results in a contradiction not only with our expectation, but also with results of measurement the Little-Parks oscillations of the resistance, obtained on the same rings. 

\subsection{The contradiction between the measurement results of the critical current and resistance oscillations.}

It was written above that our investigation have shown that the Little-Parks oscillations of resistance $R(\Phi/\Phi_{0})$ measured on the same asymmetric rings have shown that their minimums are observed at $\Phi  = n\Phi _{0}$ and their maximums are observed at $\Phi  = (n+0.5)\Phi _{0}$, Fig. 5, as well as for symmetric rings. Thus, according to these results and their universally recognized interpretation $\Delta R(\Phi/\Phi_{0}) \propto v_{s}^{2}$ \cite{10} the square of the pair velocity equals zero $v_{s}^{2} = 0$ at $\Phi  = n\Phi _{0}$ and $v_{s}^{2}$ has the maximal value at $\Phi  = (n+0.5)\Phi _{0}$ both in symmetric and asymmetric rings. According to conventional mathematics these results mean that the absolute value of the velocity $\vert v_{s}\vert  = 0$ at $\Phi  = n\Phi _{0}$ and has a maximal value at $\Phi  = (n+0.5)\Phi _{0}$ in accordance with the quantization prediction (2). The periodicity of the critical current dependencies $I_{c+}(\Phi /\Phi_{0})$, $I_{c-}(\Phi /\Phi_{0})$ with the period equal to the flux quantum can not leave doubt in that, it is quantum phenomenon caused by quantization of pair velocity in the ring (2). It is impossible to explain this periodic dependence except for the change of quantum number $n$ with change of magnetic flux $\Phi $. Therefore the position of the extreme values of the $I_{c+}(\Phi /\Phi_{0})$, $I_{c-}(\Phi /\Phi_{0})$ should, according to conventional logic, coincide with the position of the extreme values of the Little-Parks oscillations $R(\Phi/\Phi_{0})$. This coincidence is observed for symmetric ring: the maximums and minimums of the critical current $I_{c+}(\Phi /\Phi_{0})$, $I_{c-}(\Phi /\Phi_{0})$ are observed at $\Phi  = n\Phi _{0}$ and $\Phi  = (n+0.5)\Phi _{0}$ in accordance with (2). But the shift of the argument of $I_{c+}(\Phi /\Phi_{0})$, $I_{c-}(\Phi /\Phi_{0})$ on $\Delta \phi/2 = \pm 0.25$ with appearance of the ring asymmetry results to the contradiction between results of the critical current and resistance oscillations: the maximums and minimums of the $I_{c+}(\Phi /\Phi_{0})$, $I_{c-}(\Phi /\Phi_{0})$ are observed at $\Phi  = (n+0.25)\Phi _{0}$ and $\Phi  = (n+0.75)\Phi _{0}$ whereas the maximums and minimums of the resistance $R(\Phi/\Phi_{0})$ are observed as usual at $\Phi  = n\Phi _{0}$ and $\Phi  = (n+0.5)\Phi _{0}$. 

\begin{figure}
\includegraphics{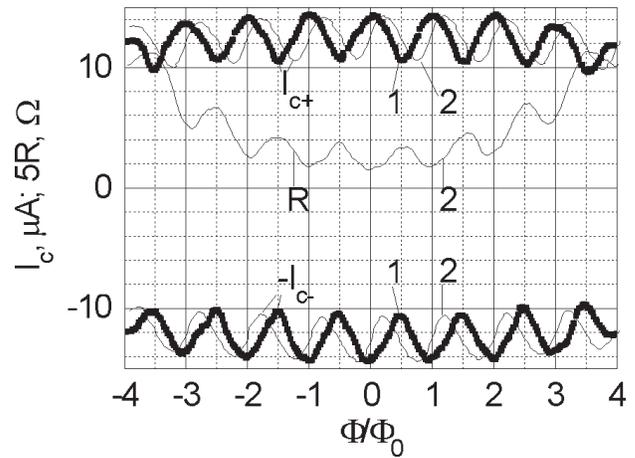}
\caption{\label{fig:epsart}   The quantum oscillations of the critical  current $I_{c+}$ $I_{c-}$ at $T = 0.975 T_{c}$, and the resistance $R$ at $T = 0.998 T_{c}$ measured on 1) the symmetric ring with $w_{n} = w_{w} = 0.4 \ \mu m$ and the asymmetric ring with $w_{n} = 0.2 \ \mu m$,  $w_{w} = 0.4 \ \mu m$.}
\end{figure}

This contradiction between the results of measurements made on the same asymmetric superconducting rings is both obvious as experimental result and inexplicable according to conventional logic. It is obvious that all periodic dependencies, both $R(\Phi/\Phi_{0})$ and $I_{c+}(\Phi /\Phi_{0})$, $I_{c-}(\Phi /\Phi_{0})$, observed on both symmetric and asymmetric rings should be functions of the argument $(n -\Phi/\Phi_{0})$. The change of this argument of the $I_{c+}(\Phi /\Phi_{0})$, $I_{c-}(\Phi /\Phi_{0})$ functions with appearance of the ring asymmetry means that either the full flux $\Phi = \Phi_{ext} + \Phi_{I}$ in the asymmetric ring differs from the one $\Phi_{ext}$ we measure or the quantum number $n$ has ceased to be an integer. Our experimental results prove that the first explanation is impossible. The additional flux $\Phi_{Iext} = LI_{ext} (s_{w}-s_{n})/2(s_{w}+s_{n})$, created by measuring current $I_{ext} < max I_{c} = 30 \ \mu A$ in our ring with inductance $L = 2 \ 10^{-11} \ H$ does not exceed $0.04\Phi_{0}$, at the maximal anisotropy $s_{w}/s_{n} = 2$. Its value $\Phi_{Iext} = LI_{ext} (s_{w}-s_{n})/2(s_{w}+s_{n})$ should depend on values of both anisotropy $s_{w}/s_{n}$ and the measuring current $I_{ext}$ which is equal to the critical current $I_{c}(T)$, at the measurement of the $I_{c+}(\Phi /\Phi_{0})$, $I_{c-}(\Phi /\Phi_{0})$ oscillations. Therefore the observations of identical shift $\pm 0.25\Phi_{0}$ at the anisotropy $s_{w}/s_{n}$ from 1.25 ($(s_{w}-s_{n})/2(s_{w}+s_{n}) = 1/9$) to 2 ($(s_{w}-s_{n})/2(s_{w}+s_{n}) = 1/3$) and critical current $I_{c}$ from $3 \ \mu A$ to $30 \ \mu A$ exclude the explanation of this shift as a consequence of the difference of the full flux in ring $\Phi $ from the measured one $\Phi _{ext}$. We can not also assume that the quantum number in the Bohr quantization can be $n \pm 0.25$ at the measurement of the critical current oscillations $I_{c+}(\Phi /\Phi_{0})$, $I_{c-}(\Phi /\Phi_{0})$ and $n$ at the measurement of the Little-Parks oscillations $R(\Phi/\Phi_{0})$.

According to the principle of realism, from two results of measurements contradicting each other, at least, one is incorrect. We observed the contradiction between the results of the measurements of $R(\Phi/\Phi_{0})$ and $I_{c+}(\Phi /\Phi_{0})$, $I_{c-}(\Phi /\Phi_{0})$ obtained on asymmetric rings with different degree of asymmetry. Our results testify that the contradiction appears only with the appearance of a ring asymmetry. It is hardly probable that these results can be incorrect. Therefore it is possible to assume that the observed contradiction is a display of violation of the principle of realism on mesoscopic level. 

\section{Conclusion. }
In order to a superconductor quantum register could be possible macroscopic quantum superposition and the EPR correlation should be observed on the mesoscopic level. Some authors \cite{25} have claimed already on observation of quantum superposition of macroscopic states in superconducting loops. But before to state on this superposition it was necessary to be convinced that single and multiple measurements give that they should give. Such investigations were made in our work. They have shown that multiply measurements do not contradict a possibility of two-states superposition whereas single measurements do not confirm existence of the two states assumed in \cite{25}. The contradiction found in our work testifies that the quantum phenomena on mesoscopic level are less clear and investigated, than it is accepted to think, proceeding from formal generalization of the quantum principles, developed for the level of elementary particles, on mesoscopic and macroscopic levels. In additional to the obvious contradiction with macroscopic realism \cite{6}, some principles of quantum mechanics come into collision with other fundamental bases of physics at these levels \cite{26}. Therefore, before to outline a way of creation of quantum computer on the mesoscopic level we should investigate comprehensively the validity and features of application of quantum principles on this level. The obvious logical contradictions and experimental results do not allow us simply to generalize the quantum formalism, which is enormously successful in describing nature at the atomic level, on higher levels. A neglect of the logical and experimental difficulties arising on the mesoscopic level may mislead about a possibility and a way of creation of quantum computer on this level.

The absence of an evidence of two state in asymmetric superconducting ring at the single measurement, found in our work, puts under doubt the statements \cite{25} on observation of quantum superposition of macroscopic states. Nobody can state about superposition of two states till he will not prove that single measurements give result corresponding to one of the permitted states. Many authors consider the existence of two permitted states in superconducting loop with half of the flux quantum as a self-evident fact. But nobody must be sure of anything in the quantum world till unambiguous experimental evidence. Einstein, Podolsky and Rosen were sure \cite{3} that a process of measurement carried out on an one system can not affect other system in any way. But experimental results \cite{5} have shown that it can, i.e. that the EPR correlation is possible. 

The undermining of the belief in the experimental evidence \cite{25} of macroscopic quantum superposition is an undermining of the belief in possibility of superconductor quantum register. But on the other hand our results inspire certain optimism concerning an opportunity of creation of a real quantum computer. There is important to realize that no quantum computer is possible without quantum superposition and the EPR correlation violating the principle of realism, i.e. a superconductor quantum register is not possible without an experimental evidence of violation of the principle of realism in a superconductor structure. This problem is not investigated for the present. The contradiction, found in our work, between the results of measurements made on the same ring may be very important for the problem of superconductor quantum register if it is indeed an experimental evidence of violation of the principle of realism on the mesoscopic level. In order to corroborate or to refute this assumption additional investigations are needed.

\section*{Acknowledgments} 
This work has been supported by a grant "Quantum bit on base of micro- and nano-structures with metal conductivity" of the Program "Technology Basis of New Computing Methods" of ITCS department of RAS, a grant of the Program "Low-Dimensional Quantum Structures" of the Presidium of Russian Academy of Sciences and a grant 04-02-17068 of the Russian Foundation of Basic Research.


\begin{thebibliography}{99}

\bibitem{1} G. Greenstein and A.G. Zajonc, \textit{ The Quantum Challenge. Modern Research on the Foundation of Quantum Mechanics.} Second Edition. Jones and Bartlett Publishers, Sudbury, Massachusetts, 2005.

\bibitem{2} A.J.Leggett, Does the everyday world really obey quantum mechanics? {\it the Public lecture at the conference "Frontiers of Quantum and Mesoscopic Thermodynamics"} 26-29 July 2004, Prague; Testing the limits of quantum mechanics: motivation, state of play, prospects. \textit{ J.Phys.: Condens. Matter} \textbf{ 14}, R415-R451 (2002); The Significance of the MQC Experiment.\textit{ J. of Superconductivity} \textbf{ 12}, 683-687 (1999).

\bibitem{3}A. Einstein, B. Podolsky, and N. Rosen,  Can Quantum-Mechanical Description of Physical Reality Be Considered Complete?  \textit{Phys. Rev.} \textbf{47}, 777 (1935). 

\bibitem{4}  J. S. Bell, On the Einstein-Podolsky-Rosen paradox. \textit{ Physics} \textbf{ 1}, 195, (1964).

\bibitem{5} A. Aspect, P. Grangier, and G. Roger, Experimental tests of realistic local theories via Bell's theorem. \textit{ Phys. Rev. Lett.} {\bf47} 460 (1981); P. G. Kwiat, K. Mattle, H. Weinfurter, and A. Zeilinger, New High-Intensity Source of Polarization-Entangled Photon Pairs. \textit{ Phys. Rev. Lett.}\textbf{ 75}, 4337 (1995); W. Tittel, J. Brendel, B. Gisin, T. Herzog, H. Zbinden, and N. Gisin, Experimental demonstration of quantum correlation over more than 10 km. \textit{ Phys. Rev. A} \textbf{ 57}, 3229 (1998); G. Weihs, T. Jennewein, C. Simon, H. Weinfurter, and A. Zeilinger, Violation of Bell's inequality under strict Einstein locality conditions. \textit{ Phys. Rev. Lett.} \textbf{ 81}, 5039 (1998).

\bibitem{6}  A. J. Leggett and Anupam Garg, Quantum mechanics versus macroscopic realism: Is the flux there when nobody looks? \textit{ Phys. Rev. Lett.} \textbf{ 54}, 857 (1985). 

\bibitem{7} N.D. Mermin, Is the moon there when nobody looks? Reality and the quantum theory. \textit{ Physics Today} \textbf{ 38}, 38-47 (1985). 

\bibitem{8} K. A. Valiev, A. A. Kokin. {\it Quantum computers: reliance and reality}, Moscow-Izhevsk: R and C Dynamics, 2002 (in Russian).

\bibitem{9} K.A.Valiev, Quantum computers and quantum computation. {\it Uspekhi Fizicheskikh Nauk} \textbf{175}, 3 (2005). 

\bibitem{10}  M.Tinkham,\textit{ Introduction to Superconductivity.} McGraw-Hill Book Company, 1975. 

\bibitem{11} Y. Makhlin, G. Schoen, and A. Shnirman, Quantum-state engineering with Josephson-junction devices. \textit{ Rev. Mod. Phys.}\textbf{ 73}, 357 (2001)

\bibitem{12} V.V. Aristov and A.V. Nikulov, Chain of superconducting loops as a possible quantum register. \textit{Proceedings SPIE} Vol.\textbf{5833}, \textit{Quantum Informatics 2004}, Yuri I. Ozhigov; Ed. pp. 145-156 (2005).

\bibitem{13} N. D. Mermin, Hidden variables and the two theorems of John Bell. \textit{ Rev. Mod. Phys.} \textbf{ 65}, 803 (1993) 

\bibitem{14} Antonio Barone, Gianfranco Paterno, \textit{ Physics and Applications of the Josephson Effect.} A Wiley-Interscience Publication, John Wiley and Sons, New York, 1981. 

\bibitem{15} J. M. Blatt, Persistent Ring Currents in an Ideal Bose Gas. \textit{ Phys. Rev. Lett.} \textbf{7}, 82 (1961); T. I. Smith, Observation of Persistent Currents in a Superconducting Circuit Containing a Josephson Junction. \textit{ Phys. Rev. Lett.} \textbf{ 15}, 460 (1965). 

\bibitem{16} L. P. Levy, et al., Magnetization of mesoscopic copper rings: Evidence for persistent currents. \textit {Phys. Rev. Lett.} \textbf{64}, 2074 (1990); D. Mailly, C. Chapelier, and A. Benoit, Experimental observation of persistent currents in GaAs-AlGaAs single loop. \textit{ Phys. Rev. Lett.} \textbf{ 70}, 2020 (1993).

\bibitem{17}F. Balestro, J. Claudon, J. P. Pekola, O. Buisson, Evidence of two-dimensional macroscopic quantum tunneling of a current-biased DC-SQUID. \textit{ Phys. Rev. Lett.} \textbf{ 91}, 158301 (2003); Shao-Xiong Li, Yang Yu, Yu Zhang, Wei Qiu, Siyuan Han, and Zhen Wang, Quantitative Study of Macroscopic Quantum Tunneling in a dc SQUID: A System with Two Degrees of Freedom. \textit{ Phys. Rev. Lett.} \textbf{89}, 098301 (2002). 

\bibitem{18} A.V. Nikulov, Quantum force in superconductor. \textit{ Phys. Rev. B} \textbf{ 64}, 012505 (2001).

\bibitem{19} H.Vloeberghs, V.V.Moshchalkov, C. Van Haesendonck, R.Jonckheere, and Y.Bruynseraede, Anomalous Little-Parks Oscillations in Mesoscopic Loops. \textit{Phys. Rev. Lett.} \textbf{69}, 1268 (1992).

\bibitem{20} W.A.Little and R.D.Parks, Observation of Quantum Periodicity in the Transition Temperature of a Superconducting Cylinder. \textit{Phys. Rev. Lett.} \textbf{9}, 9 (1962).

\bibitem{21} I.N.Zhilyaev, S.G.Boronin, A.V.Nikulov, K.Fossheim, States in the 
structure of weakly connected superconducting rings. \textit{Quantum Computers and Computing} \textbf{2}, 49 (2001).

\bibitem{22} D. S. Golubovic and V. V. Moshchalkov, Linear magnetic flux amplifier. \textit{Appl. Phys. Lett.} \textbf{ 87}, 142501 (2005).

\bibitem{23} S.V.Dubonos, V.I.Kuznetsov, I.N.Zhilyaev, A.V.Nikulov, and A.A.Firsov, Induction of dc voltage, proportional to the persistent current, by external ac current on system of inhomogeneous superconducting loops. {\it Pisma Zh. Eksp. Teor. Fiz} {\bf 77}, 439 - 444, (2003) ({\it JETP Lett.} {\bf 77}, 371 - 375, 2003); e-print arXiv: cond-mat/0303538

\bibitem{24} A.Th. A.M. De Waele, W.H.Kraan, R. De Bruynouboter and K.W. Taconis, On the D.C. Voltage across a Double Point Contact between Two Superconductors at Zero Applied D.C. Current in Situations in Which the Junction is in the Resistive Region due to the Circulating Current Flux Quantization. \textit{Physica} \textbf{ 37}, 114 (1967).

\bibitem{25} J. R. Friedman et al, Quantum superposition of distinct macroscopic states. \textit{Nature} \textbf{ 406}, 43 (2000); C.H. van der Wal et al., Quantum superposition of macroscopic persistent-current states. Science \textbf{290}, 773 (2000). 

\bibitem{26} A.V. Nikulov, How a Possibility of the Einstein-Podolsky-Rosen Correlation on Mesoscopic Level Could be Investigated. \textit{the talk at the 13th General Meeting of the European Physical Society "Beyond Einstein - Physics for the 21th Century",} Bern, Switzerland, 11-15 July 2005; e-print arXiv: cond-mat/0506653.

\end{thebibliography}
\end{document}